\documentclass{ieeeaccess}
\usepackage{cite}
\usepackage{amsmath,amssymb,amsfonts}
\usepackage{algorithmic}
\usepackage{graphicx}
\usepackage{caption}
\usepackage{hyperref}
\usepackage[caption=false,font=footnotesize]{subfig} % IEEEtran-friendly subfigures
\usepackage{textcomp}
\usepackage{booktabs}
\def\BibTeX{{\rm B\kern-.05em{\sc i\kern-.025em b}\kern-.08em
    T\kern-.1667em\lower.7ex\hbox{E}\kern-.125emX}}
\begin{document}
\history{Date of publication xxxx 00, 0000, date of current version xxxx 00, 0000.}
\doi{10.1109/ACCESS.2017.DOI}

\title{Assessing Band Gap Stability of Organic Semiconductor Thin Films for Flexible Electronic Applications}
\author{\uppercase{Mahya Ghorab}\authorrefmark{1},
\uppercase{Ayush K. Ranga\authorrefmark{2}, Arnulf Materny\authorrefmark{2},
Veit Wagner}.\authorrefmark{2},
, \uppercase{Mojtaba Joodaki}\authorrefmark{1}\IEEEmembership{Senior Member, IEEE}}
\address[1]{School of Computer Science \& Engineering, Constructor University, Bremen, Germany}
\address[2]{School of Science, Constructor University, Bremen, Germany}

\markboth
{Author \headeretal: Preparation of Papers for IEEE TRANSACTIONS and JOURNALS}
{Author \headeretal: Preparation of Papers for IEEE TRANSACTIONS and JOURNALS}

\corresp{Corresponding author: Mojtaba Joodaki (e-mail: mjoodaki@constructor.university).}

\begin{abstract}
Integration of organic semiconductors into flexible electronics requires that their optoelectronic properties remain stable under mechanical deformation. Among these, the optical band gap governs exciton generation and limits photovoltaic voltage, making it a key parameter for strain-resilient design. In this work, we investigate band gap shifts in poly(3-hexylthiophene-2,5-diyl) (P3HT) and poly(3,4-ethylenedioxythiophene):poly(styrenesulfonate) (PEDOT:PSS)/P3HT  thin films deposited on flexible poly(ethylene terephthalate) (PET) substrates under uniaxial tensile strain ranging from 1\% to 10\%. Samples were subjected to mechanical deformation and then characterized by ultraviolet--visible (UV--Vis) absorption spectroscopy. The optical band gaps extracted using a standardized Tauc analysis and statistically validated through equivalence testing and robust regression models. We find that up to 7\% strain, the band gap shift ($\Delta E_g$) remains effectively invariant, independent of annealing condition or stack configuration, demonstrating electronic stability. However, at 10\% strain, all groups exhibit a reproducible widening of $\sim$4--5~meV. This threshold-like behavior marks a transition from mechanical accommodation to electronic perturbation. These findings confirm that the optical band gap in semicrystalline P3HT-based thin films is robust under practical deformation, which provides clear strain thresholds to inform mechanical modeling and device-level simulation of flexible organic optoelectronic systems.
\end{abstract}


\titlepgskip=-15pt

\maketitle

\section{Introduction}
\label{sec:introduction}
\PARstart{O}{ver} the past decade, a substantial body of research has emerged focusing on flexible and stretchable organic solar cells (OSCs). The primary goal of these studies, such as those by Wardani \textit{et al.} \cite{WARDANI}, Dauzon \textit{et al.} \cite{Dauzon2021}, and Wang \textit{et al.} \cite{Wang2018},  has been to enhance device architectures for improved mechanical compliance, durability, and application versatility in wearable and bio-integrated electronics. While this body of work demonstrates significant progress, strain is generally framed as an external constraint to be mitigated, rather than as a dynamic variable intrinsically modulating the physical behavior of device materials\cite{Menichetti2017,Kim2021,Ghorab2025,Aboutorabi2015,Aboutorabi2014,Khorami2019}. The influence of mechanical stress on the optoelectronic properties of functional layers, such as the active layer, remains only partially understood. These strain induced alterations including changes in optical absorption, molecular alignment, and charge transport pathways are critical because they ultimately determine the propagation of stress through the device and modulate its overall performance \cite{Salari2019,Salari2018,Joodaki2018,Ghorab2022}. Yet such effects remain largely uncharacterized under realistic processing and application conditions, limiting both our predictive modeling capabilities and the rational design of truly strain-resilient photovoltaic systems. \\
To properly address these knowledge gaps, it is important to briefly review prior investigations that have examined how mechanical strain impacts the optoelectronic and charge-transport properties of organic semiconductors. Early work by Sokolov \textit{et al.} \cite{Sokolov2012} focused on bending induced strain in organic field effect transistors (OFETs). The results revealed that charge carrier mobility is highly sensitive to the dielectric/semiconductor interface. These  modifications were attributed to the dielectric surface energy, which perturbs orbital overlap and transport pathways at the interface. This study provided one of the first mechanistic explanations for strain sensitivity in device level geometries, though it did not directly address optoelectronic properties such as band gap or absorption edge stability. \\
Subsequent studies expanded this scope to molecular semiconductors such as rubrene. Using scanning Kelvin probe microscopy and in-situ X-ray diffraction, Wu \textit{et al.} \cite{Wu2016} demonstrated that the work function (WF) of rubrene single crystals responds strongly to both tensile and compressive strain. Their findings, supported by density functional theory (DFT) calculations, indicated that the work function increases under tensile strain and decreases under compressive strain. In parallel, Kubo \textit{et al.} \cite{Kubo2016} examined uniaxial tensile strain in organic semiconductors and observed anisotropic degradation of charge transport. They related this behavior to direction-dependent changes in intermolecular overlap and to enhanced dynamic disorder. These observations suggest that strain does not affect transport in a uniform manner but instead depends on both the axis of deformation and the packing orientation of the molecules. Together, these studies provided some of the first evidence that mechanical strain can alter essential electronic parameters in organic semiconductors. Most of the mentioned works, however, were carried out on single crystals or highly ordered thin films. Furthermore, while strain induced perturbations to molecular orbitals and charge transport have been demonstrated, their direct manifestation in macroscopic observables such as absorption edge shifts or photovoltaic performance metrics remains largely unexplored.\\
Building on these works, more efforts have shifted from small molecule crystals to polymeric semiconductors, where mechanical compliance and semicrystalline disorder create qualitatively different strain responses. Recent theoretical and ultrafast imaging studies have suggested that even modest strain can couple directly to the electronic structure of conjugated polymers, potentially altering band alignment or excitonic processes in well-characterized conjugated polymers such as poly(3-hexylthiophene) (P3HT) \cite{Menichetti2017,Kim2021}. While these insights point to the possibility that mechanical deformation could actively modulate optoelectronic properties, it remains unclear whether such molecular  and nanoscale effects persist in thin films under realistic processing and operating conditions. This uncertainty motivates the present study, where we experimentally probe the stability of the optical band gap in strained P3HT thin films.\\
While prior studies demonstrate that strain can influence the electronic behavior of organic semiconductors, they do not yet clarify how such effects translate to technologically relevant thin films under practical deformation. A key open question is the stability of the optical band gap, a parameter that defines light absorption, governs exciton generation at the molecular scale, and ultimately dictates device level outputs such as open circuit voltage ($V_{oc}$), short circuit current ($I_{sc}$), fill factor (FF), and power conversion efficiency (PCE) \cite{Ghorab2022,Joodaki2018,Salari2018, Ghorabreview,Khoshdel2019}. Despite its central importance, the response of the band gap to mechanical strain remains only partially characterized.\\ 
In this work, we investigate whether residual mechanical strain in the range of 0--10\% alters the optical band gap of P3HT thin films deposited on flexible poly(ethylene terephthalate) (PET) substrates, both as single layer and in poly(3,4-ethylenedioxythiophene):poly(styrenesulfonate) (PEDOT:PSS)
/P3HT stacks. In practical applications such as foldable displays or wearable electronics, the tensile strain experienced by polymer substrates and electrode stacks is typically only a few percent. Values in the range of 1--5\% are most commonly reported as the limit for reliable operation before mechanical degradation sets in~\cite{Saleh2021}. The 1--7\% window investigated in this study therefore encompasses and slightly exceeds realistic device conditions. By contrast, the 10\% condition represents a deliberately severe deformation, included here to reveal the strain threshold at which measurable electronic changes begin to appear. To this end, we applied controlled tensile deformation using a custom-built stretcher and performed ultraviolet--visible (UV--Vis) absorption spectroscopy before and after stretching, extracting the band gap through a standardized Tauc analysis. This approach allows us to directly test the stability of the absorption edge in a representative layer stack for OSCs, providing insight into how mechanical deformation translates into optoelectronic resilience under conditions relevant to long-term device operation.

\section{Experimental Setup and Methods}
\subsection{Sample Preparation}
% Requires the same packages as above
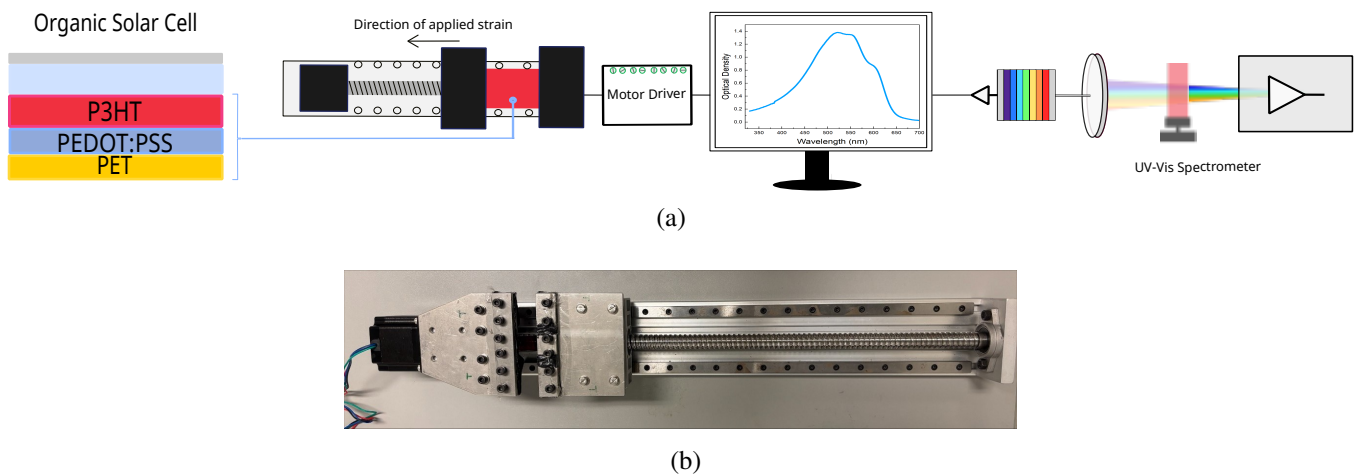
\begin{figure*}[t]
  \centering

  % Panel A (top)
  \begin{minipage}{\textwidth}
    \centering
    \begin{picture}(0,0)
      \put(-9,-80){(a)} % adjust coords if needed
    \end{picture}
    \includegraphics[width=\textwidth]{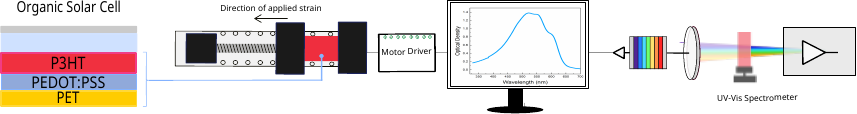}
  \end{minipage}

  \vspace{3em}

  % Panel B (bottom)
  \begin{minipage}{\textwidth}
    \centering
    \begin{picture}(0,0)
      \put(125,-15){(b)} % adjust coords if needed
    \end{picture}
    \includegraphics[width=0.5\textwidth]{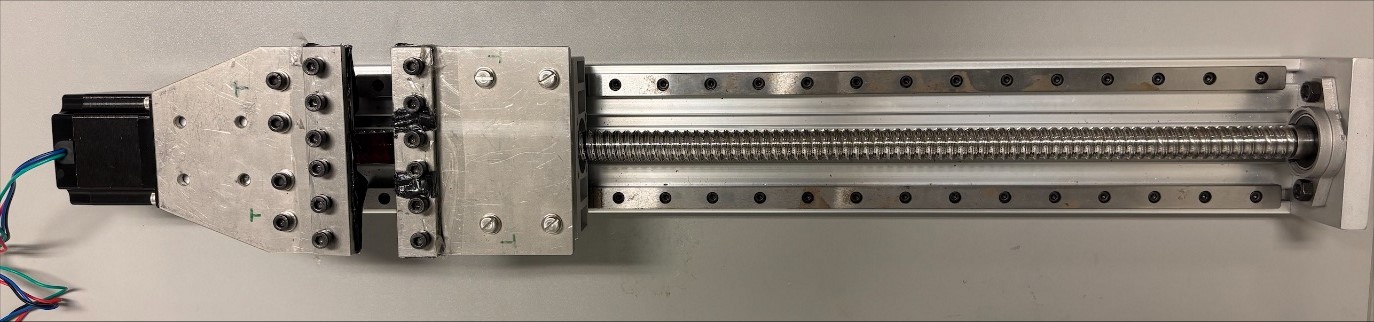}
    \vspace{3em}
  \end{minipage}

  \caption{Two-panel overview of the measurement workflow and apparatus.
  \textbf{(a)} Schematic illustration of the experimental workflow. Samples were first characterized by UV–Vis spectroscopy in their unstretched state, then subjected to tensile strain for 30 minutes using a motorized stretcher, and subsequently transferred back to the spectrometer for a second UV–Vis measurement. 
  \textbf{(b)} Photograph of the custom motorized stretcher used to apply strain. 
  Note that the mechanical straining and optical measurements were performed sequentially (ex-situ).}
  \label{fig:uvvis-detector-wide}
\end{figure*}

Regioregular P3HT with an average molecular weight of 20,000--45,000 g/mol from Sigma-Aldrich was used as the donor polymer. Flexible PET with a thickness of 57.3 $\mu$m was sourced from a roll, which was manufactured for OSC applications, characterized by high optical clarity and an amorphous morphology \cite{Ghorab2025}. For experimental use, PET sheets were cut into $5 \times 5$ cm$^{2}$ pieces, cleaned sequentially with a 3:2 mixture of propanol and acetone, and dried with a nitrogen gun. To improve surface wettability and adhesion, the substrates were further treated with UV--ozone (UV/O$_3$) treatment
for 5 minutes in a sealed chamber immediately before film deposition. \\
A 33 mg/mL solution of P3HT in anhydrous chlorobenzene was prepared and stirred overnight at 50$^{\circ}$C to ensure complete dissolution. The PET substrates were pre-heated at 75$^{\circ}$C for 10 minutes prior to deposition. Films were deposited by spin-coating 400 $\mu$L of solution at 1500 rpm for 30 seconds. After the deposition, the films were divided into three groups: (i) no thermal treatment (unannealed), (ii) annealing at 50$^{\circ}$C for 5 minutes, and (iii) annealing at 75$^{\circ}$C for 5 minutes. \\
In addition to these single layer samples, bilayer structures incorporating a thin hole transport layer (HTL) of PEDOT:PSS were also fabricated. In this context, PEDOT:PSS was employed primarily to enhance interfacial adhesion, as it provides excellent mechanical coupling between PET and P3HT during subsequent stretching experiments. PEDOT:PSS
 (formulation containing 98\% PEDOT and 2\% surfactant additive) was spin coated onto PET substrates at 5000 rpm for 60 s, using 1000\,$\mu$L of dispersion per sample. The films were annealed at 130$^{\circ}$C for 30 minutes to remove residual solvent. The P3HT layer was then deposited on top under the same spin coating conditions as described above, yielding PET/PEDOT:PSS/P3HT structure with improved mechanical stability.  
\subsection{Experimental Setup}
After annealing, test specimens with dimensions of 15 mm $\times$ 33 mm were cut from the larger substrates. The effective gauge length between clamps was 25 mm, with 4 mm on each side reserved for clamping. Uniaxial tensile strain was applied using a custom-built mechanical stretcher based on a precision linear actuator. Calibration confirmed a displacement resolution of 6.25 $\mu$m per microstep, allowing reproducible strain increments.
The maximum strain applied in this study was limited to 10\%. Beyond this point, PET substrates undergo pronounced mechanical distortion, which would dominate the spectral response and obscure any polymer specific effects \cite{Ghorab2025}. In this study, strain values of 1\%, 3\%, 5\%, 7\% and 10\% were applied. The strain ($\varepsilon$) was calculated as the ratio of the applied elongation $\Delta L$ to the initial gauge length $L_0$  ($\varepsilon = \Delta L / L_0$). All stretching and optical measurements were performed under ambient room temperature conditions, ensuring that no softening or morphological changes associated with elevated temperatures influenced the results.\\
UV--Vis absorption spectra were recorded using a PerkinElmer Lambda~12 spectrophotometer in the 200--1000 nm wavelength range with a spectral resolution of 1 nm. Measurements were performed under ambient conditions with unpolarized light. For all films, three spectra were acquired per sample and averaged to improve signal stability. Identical protocols were used across all sample types.\\
To ensure that the reported absorption spectra reflect only the active organic layers, the optical contribution of the PET substrate was explicitly accounted for. In a previous study, we characterized the strain response of bare PET under the same tensile conditions used here \cite{Ghorab2025}. That dataset was used to subtract the PET baseline from all spectra collected in this work. As a result, the reported absorption and extracted band gaps correspond solely to P3HT or PEDOT:PSS/P3HT films, with any PET-related absorption or strain effects removed.  \\
Strain application in this work followed a residual strain protocol. Each sample was first measured in its pristine state, then subjected to a fixed tensile strain for 30 minutes using the stretcher. After this period, the strain was released and the film was immediately transferred back to the spectrometer for re-measurement. Importantly, no relaxation time was allowed between strain release and optical characterization, ensuring that the spectra captured the post-deformation state of the film. To avoid artifacts from strain history, a fresh film was used for each strain level. Fig.\ref{fig:uvvis-detector-wide}, depicts the schematic of this setup. In this study, ``residual strain'' refers to structural modifications in the organic layer that persist after unloading, rather than a direct measurement of sample length before and after stretching. This approach provides a direct measure of morphological stability which are the irreversible or long-lived changes that remain once mechanical stress is released. From a device perspective, this is particularly relevant since operational reliability depends not only on transient behavior under load, but also on whether the active layers recover or drift structurally once the strain is removed, before the next cycle of use. While transient in-situ electronic changes during active deformation are also of fundamental interest, they are beyond the scope of the present study and could be addressed in future in-situ spectroscopy experiments.\\

\subsection{Tauc Analysis and Optical band gap Extraction}

The electronic band gap in organic semiconductors can be characterized through multiple, complementary approaches. The fundamental gap, defined as the energy difference between the valence band maximum (VBM) and conduction band minimum (CBM), represents the intrinsic electronic structure of the material. This quantity can be accessed via first-principles band-structure calculations, or experimentally through the combination of photoelectron spectroscopy (PES) and inverse photoelectron spectroscopy (IPES), which directly probe the ionization and electron affinity levels, respectively \cite{Huang2014}. In contrast, electrochemical techniques such as cyclic voltammetry (CV) produce the transport gap, corresponding to the energy required to generate free charge carriers, defined by the difference between the ionization potential (IP) and electron affinity (EA) \cite{Bredas2009}.  \\
In conjugated polymers, optical excitation typically generates bound excitons rather than free charges, due to the low dielectric screening and strong Coulomb interactions \cite{Dimitriev2022}. As a result, the optical gap, which dictates the absorption onset, is systematically smaller than the transport gap by the exciton binding energy. This optical transition plays a central role in defining the spectral window for photon harvesting and constrains the achievable $V_{oc}$ in organic photovoltaics \cite{Scharber2013}.  Optical properties of P3HT can by deduced in reflection geometry by
ellipsometry or modulation spectroscopy \cite{Pittner2013}, or more convenient in
transmission by UV-Vis absorption. However, precise extraction of the optical gap from UV–Vis absorption spectra is often hindered by vibronic fine structure and inhomogeneous broadening arising from energetic disorder \cite{Spano2010}.  \\
To determine the optical band gap, we employed the Tauc method \cite{Tauc1966,Yu2010}, which relates the absorption coefficient $\alpha$ to photon energy $h\nu$ as:
\begin{equation}
(\alpha h\nu)^\frac{1}{n} = B(h\nu - E_g),
\end{equation}
where $\alpha$ is the absorption coefficient, $h$ is Planck's constant, $\nu$ is the photon frequency, $E_g$ is the optical band gap, $B$ is a proportionality constant, and $n$ is the transition exponent that depends on the nature of the electronic transition ($n = \tfrac{1}{2}$ for direct allowed transitions and $n = 2$ for indirect allowed transitions) \cite{Kerremans2020,Mamand2025,Hang2023,Makula2018,Scharber2021,Zhong2023}. In this study, decadic absorbance $A$ was used in place of the absorption coefficient $\alpha$ to eliminate the influence of film thickness variations on the band gap determination. Since $\alpha = (2.303/d)A$, where $d$ is the film thickness, the use of absorbance $A$ in place of $\alpha$ merely rescales the vertical axis of the Tauc plot without altering the $x$-intercept that defines $E_g$. The nominal thicknesses of our spin-coated films were $\sim$200 nm for P3HT and $\sim$25 nm for PEDOT:PSS, values consistent with prior reports under similar deposition conditions \cite{Muller2017}. However, small sample-to-sample variations in $d$ (typically on the order of tens of nanometers) can introduce artificial scatter when calculating $\alpha$. Because all films were fabricated under identical spin-coating conditions, residual thickness variations are systematic across the dataset and do not bias the relative comparison of optical band gaps. For this reason, absorbance $A$ was used directly, as it provides a more robust and reproducible metric for $E_g$ extraction while decoupling the analysis from uncertainties associated with precise thickness determination.\\
For P3HT we adopted $n=1/2$, which is considered for direct band gap semiconductors \cite{Tsumuraya2012} and consistently provided a broader and more stable linear region near the onset, while avoiding distortions from vibronic shoulders \cite{Scharber2013}.  \\
For each specimen, three spectra were recorded, averaged, and smoothed with a Savitzky--Golay filter (polynomial degree 3). The absorbance spectrum was converted to photon energy using $h\nu~[eV]= 1240/\lambda$~[nm] , where $\lambda$ is the wavelength, and the Tauc variables were defined as $x = h\nu$ and $y = (A h\nu)^{2}$. A fixed energy interval between 1.90 and 2.026 eV was selected as the quasi-linear onset region (corresponding to $\sim$604 nm at the absorption edge of P3HT). Within this interval, a fixed number of contiguous points were always used to ensure comparability across spectra.  \\
Linear fits were performed using ordinary least squares (OLS) in centered form:
\begin{equation}
y = m(x - \bar{x}) + b_c,
\end{equation}
where $\bar{x}$ is the mean energy in the window, $m$ the slope, and $b_c$ the centered intercept. The band gap was obtained from the $x$-intercept:
\begin{equation}
E_g = \bar{x} - \frac{b_c}{m}.
\end{equation}
Uncertainty in $E_g$ was estimated by error propagation (delta method) from the regression covariance matrix of $m$ and $b_c$, using:
\begin{equation}
\frac{\partial E_g}{\partial m} = \frac{b_c}{m^2}, 
\qquad
\frac{\partial E_g}{\partial b_c} = -\frac{1}{m}.
\end{equation}
This yielded standard errors (SEs) and 95\% confidence intervals (CI) for each fit. Among all fixed-length windows within the chosen interval, the optimal fit was defined as the one with the smallest SE, provided that the coefficient of determination ($R^2$) exceeded 0.995 and the residuals showed no curvature. 
To account for instrument resolution, the 1~nm spectral resolution (\(\approx 3\text{–}5~\mathrm{meV}\) at the P3HT onset) was adopted as the detection limit. \\ %Energy differences below \(\pm 5~\mathrm{meV}\) were treated as within experimental uncertainty.
\begin{figure*}[!t]
  \centering

  % (a)
  \begin{minipage}[t]{0.31\textwidth} % <-- note the [t]
    \centering
    \includegraphics[width=\linewidth,keepaspectratio]{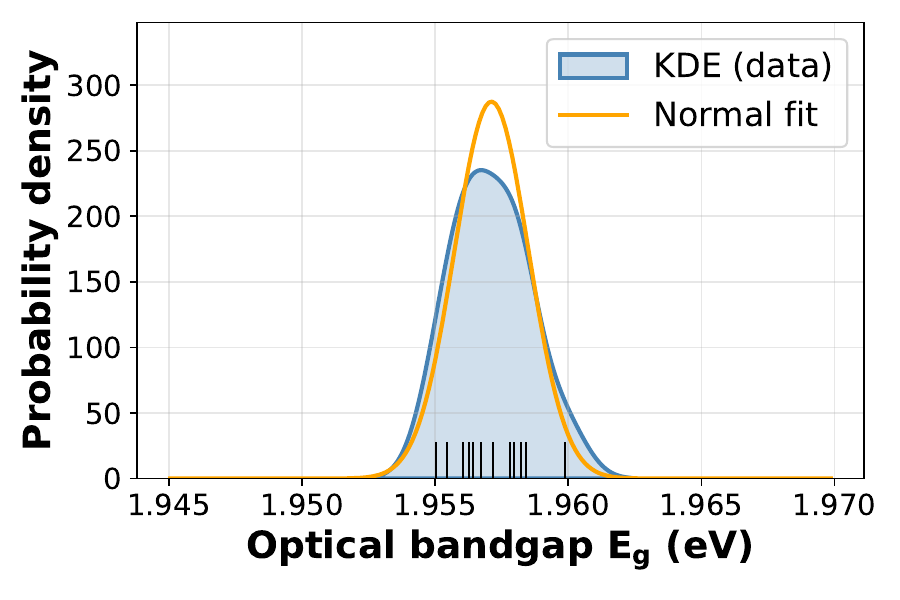}
    \vspace{2pt}
    {\footnotesize (a) Unannealed}
  \end{minipage}\hfill
  % (b)
  \begin{minipage}[t]{0.31\textwidth} % <-- note the [t]
    \centering
    \includegraphics[width=\linewidth,keepaspectratio]{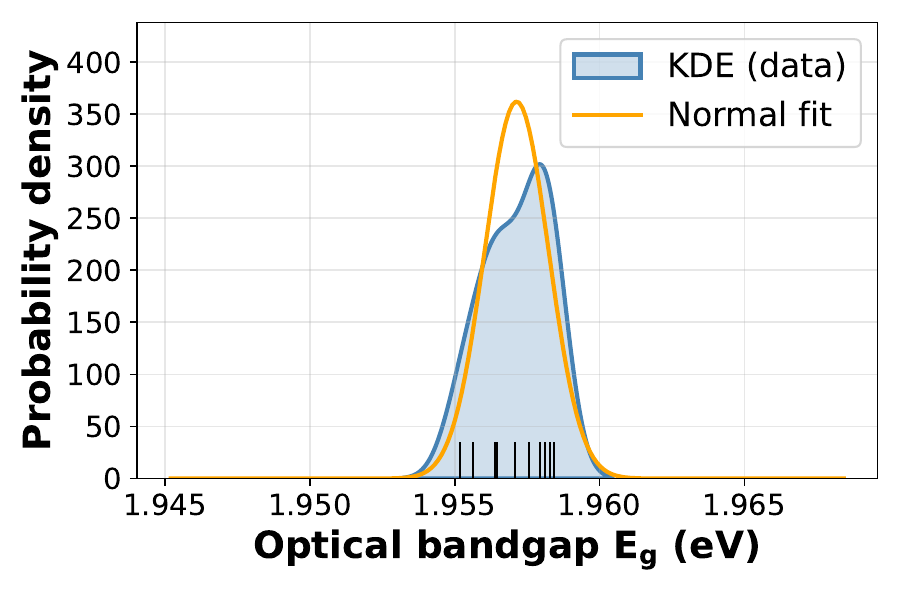}
    \vspace{2pt}
    {\footnotesize (b) 50$^{\circ}$C}
  \end{minipage}\hfill
  % (c)
  \begin{minipage}[t]{0.31\textwidth} % <-- note the [t]
    \centering
    \includegraphics[width=\linewidth,keepaspectratio]{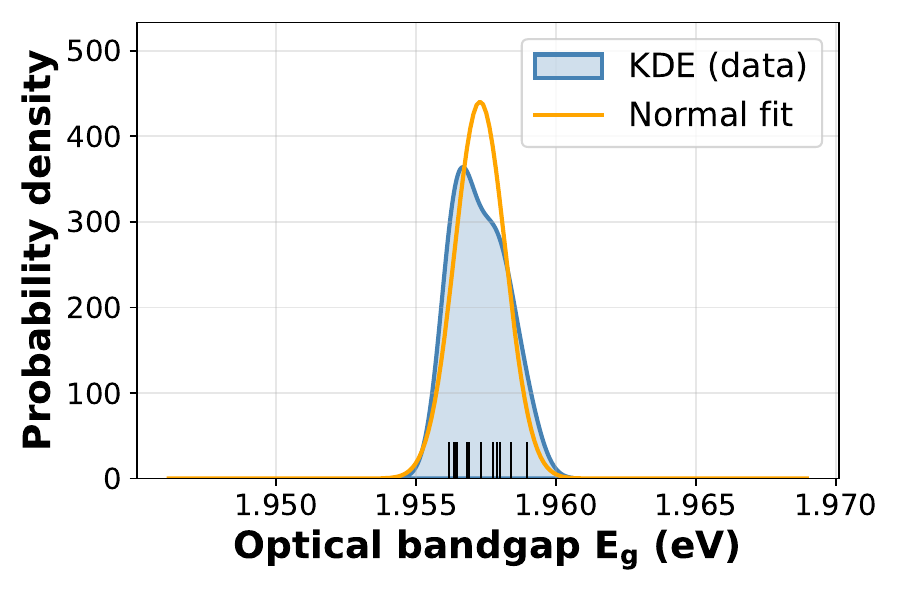}
    \vspace{2pt}
    {\footnotesize (c) 75$^{\circ}$C}
  \end{minipage}

  \caption{KDEs of the optical band gap $E_g$ under three annealing conditions, with fitted normal distributions for reference. The KDEs use a Gaussian kernel (base function) and Scott's bandwidth rule ($h = \sigma n^{-1/5},\ n=12$)~\cite{Scott1992}. Rug marks along the axis show the 12 measured $E_g$ values for each condition.}
  \label{fig:kde_all}
\end{figure*}
Twelve independent spectra were analyzed for each annealing condition (unannealed, 50\,\textdegree C, 75\,\textdegree C), reflecting the total number of experiments conducted per condition and stacking configuration. The extracted band gap values were then aggregated into condition-specific descriptor sets. For each set, the arithmetic mean band gap (\( \overline{E}_\mathrm{g} \)) was reported as the central value, with the sample standard deviation  (SD, \( \sigma \)) and variance (\( \sigma^2 \)) quantifying variability arising from sample-to-sample fabrication differences. A 95\% CI for the mean was calculated using the Student’s t-distribution, \( \overline{E}_\mathrm{g} \pm t_{0.975,n-1} \cdot \sigma/\sqrt{n} \), to capture statistical uncertainty.\\
In parallel, \( R^2 \) values from individual linear fits were averaged to provide an overall metric of fit quality. Fig. \ref{fig:kde_all} visualizes the distribution of \( E_\mathrm{g} \) values across annealing conditions, kernel density estimates (KDEs) were plotted and overlaid with fitted normal distributions. These plots reveal narrowly distributed profiles that closely follow the fitted Gaussian functions, verifying the adequacy of a single-Gaussian description of the data. The use of twelve extracted $E_g$ values per condition to construct both the KDE and the Gaussian fit reflects fabrication-induced variability, which is expected to follow a normal distribution. The small differences between KDE curves and the Gaussian fits can be attributed to the limited sample size ($n=12$), rather than to systematic deviations from normality. A summary of the extracted parameters, including mean \( E_\mathrm{g} \), SD, CI, and average \( R^2 \) is provided in Table \ref{tab:band gap_summary}.
\begin{table}[h]
\centering
\caption{Summary of optical band gap parameters for P3HT films. Values averaged over 12 spectra per condition.}
\label{tab:band gap_summary}
\begin{tabular}{lcccc}
\hline
Anneal Condition & Mean $E_g$ & SD & 95\% CI & Mean $R^2$ \\
\hline
Unannealed  & 1.9571 & 0.0013 & 1.9562 – 1.9579 & 0.997 \\
50$^{\circ}$C & 1.9571 & 0.0011 & 1.9564 – 1.9578 & 0.997 \\
75$^{\circ}$C & 1.9572 & 0.0009 & 1.9567 – 1.9578 & 0.997 \\
\hline
\end{tabular}
\end{table}
\begin{figure}[!t]
\centering
\includegraphics[width=\columnwidth]{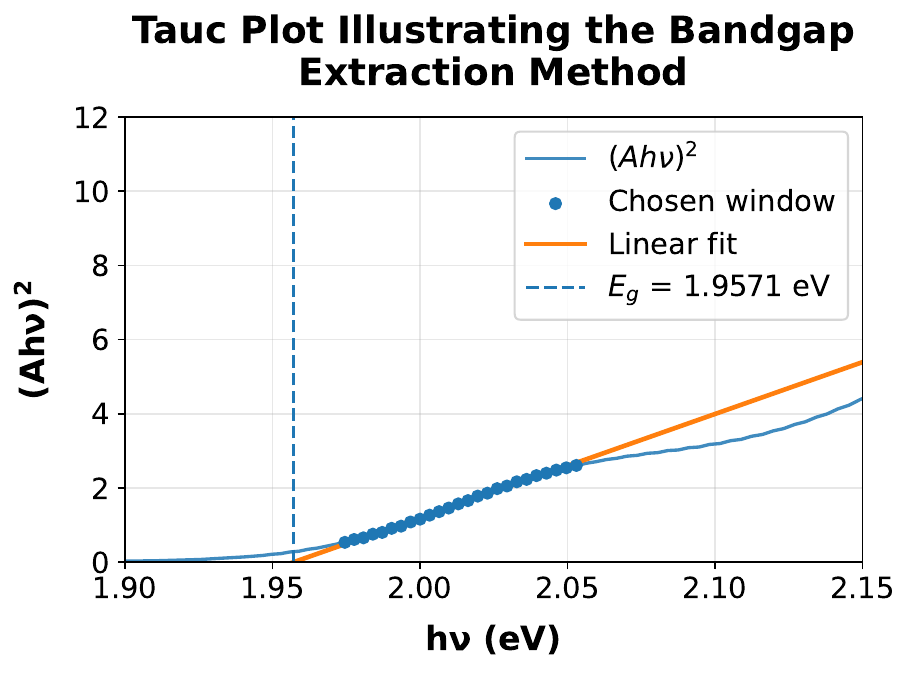}
\caption{Representative Tauc plot illustrating the method used to extract the optical band gap $E_g$. The straight line corresponds to the optimal linear fit within the selected onset region, and the dashed vertical line marks the intercept used to determine $E_g$.}
\label{fig:tauc_example}
\end{figure}\\
The baseline values of \( E_g \) for unstrained P3HT films under different annealing conditions are summarized in Table~\ref{tab:band gap_summary}. The mean values differ by less than 0.1~meV across conditions, indicating that the average optical gap remains essentially unchanged within experimental uncertainty. However, the width of the 95\% CI narrows upon annealing, decreasing from 1.7~meV in the unannealed state to 1.1~meV at 75\,$^{\circ}$C. This reduction in variability reflects improved structural order and more uniform film morphology. At the relatively low annealing temperatures applied here, the dominant effect is thus a suppression of energetic disorder rather than a pronounced shift in the mean band gap. These temperatures correspond to commonly employed fabrication protocols for P3HT-based organic solar cells on flexible substrates, where annealing is usually restricted to below $\sim$80~$^{\circ}$C.
 This observation is consistent with prior reports showing that thermal treatment enhances crystallinity in poly(3-alkylthiophenes) \cite{Prosa1992,Verploegen2010,Muller2017}. Fig. \ref{fig:tauc_example}, depicts the Tauc method plot, which is used for the band gap extraction.
 
\section{Results and Discussion}

\begin{figure}[!t]
  \centering
  \includegraphics[width=0.48\textwidth]{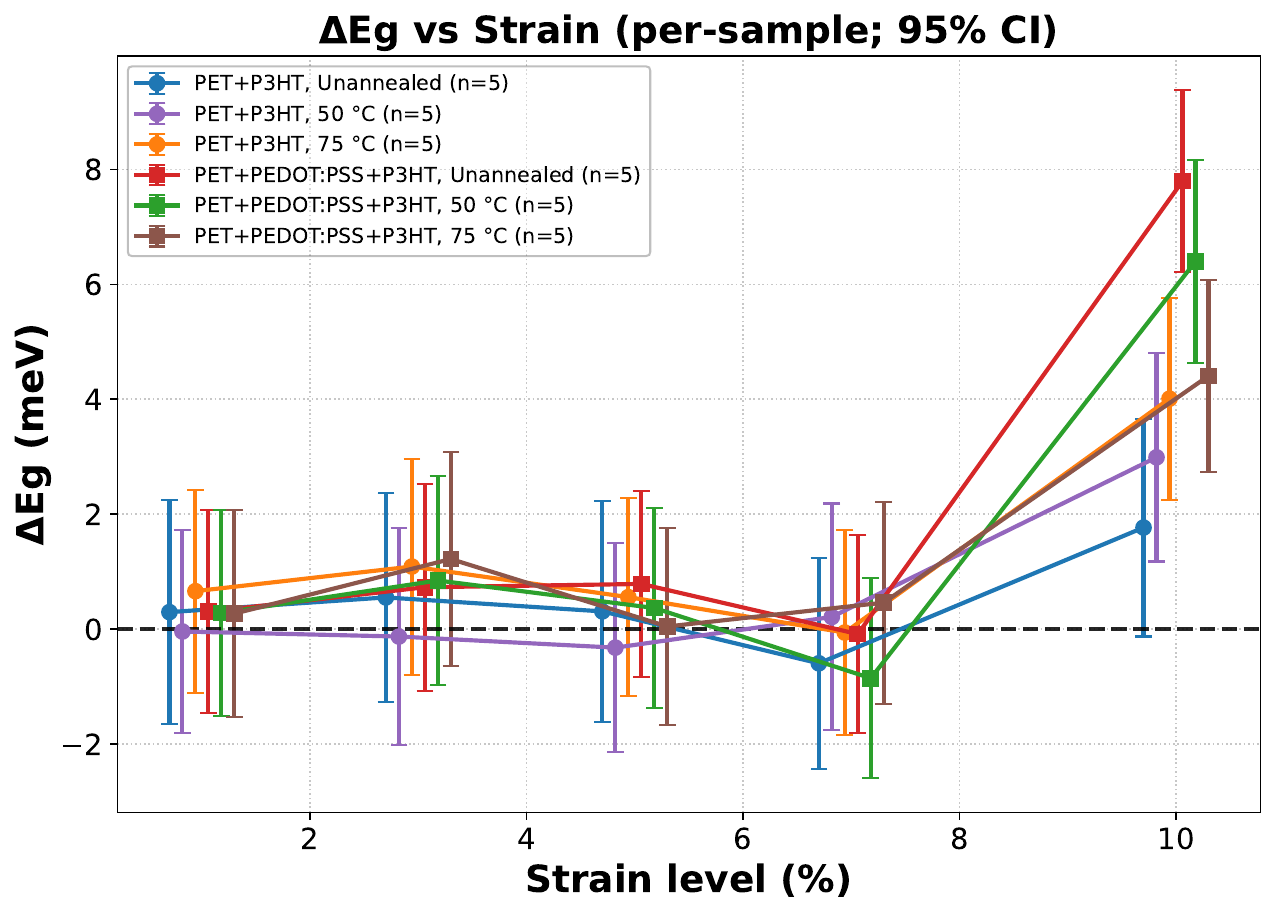}
  \caption{Mean band gap shift ($\Delta E_g$) versus strain with 95\% CIs across all stacks and annealing conditions.}
  \label{fig:deltaeg_vs_strain_overview}
\end{figure}

To access the effect of mechanical strain on the optical band gap, we calculated the per-sample difference

\[
\Delta E_g = E_{g,\mathrm{post}} - E_{g,\mathrm{pre}},
\]
where \( E_{g,\mathrm{pre}} \) and \( E_{g,\mathrm{post}} \) denote the band gaps measured before and after strain application, respectively. The uncertainties from the Tauc fits were carried through so that each 
 \( \Delta E_g \) value came with its own estimated error. The data were then grouped according to strain level, stacking configuration, and annealing temperature.\\
Across all conditions, the resulting shifts are generally small (on the order of a few meV). These values often fall within the experimental resolution limit of \( \pm 5 \,\mathrm{meV} \), which is determined by the 1\,nm spectrometer step size. Fig.~\ref{fig:deltaeg_vs_strain_overview} presents the aggregated results as \( \Delta E_g \) versus strain, with 95\% CIs. For strains up to 7\%, the mean values fluctuate around zero and remain largely confined within the resolution band which would suggest the absence of a systematic effect. However, at 10\% strain, all groups exhibit a reproducible increase in \( E_g \) of approximately 4--5~meV, which, while modest in magnitude, clearly exceeds the estimated statistical noise.\\
\begin{table*}[t]
\centering
\caption{Summary of statistical tests evaluating strain sensitivity of $\Delta E_g$.}
\label{tab:strain_tests}
\begin{tabular}{p{2.5cm} p{2cm} p{3cm} p{3.5cm} p{4.5cm}}
\toprule
Contrast & Estimate & CI & Test & Conclusion \\
\midrule
Slope (0--7\% strain) 
& $-0.089$ meV/\% 
& 90\% CI [--0.173, --0.005] 
& TOST, margin $\pm 0.5$ meV/\% 
& Effect statistically equivalent to zero; max shift $\leq 2$ meV \\
10\% vs. pooled 1--7\% 
& $+4.27$ meV 
& 95\% CI [2.38, 6.17] 
& Robust categorical regression (HC3) 
& Significant strain-induced permanent changes occur between 7 and 10\%\\
\bottomrule
\end{tabular}
\end{table*}
These observations suggest a threshold-like response in the optical band gap, which is stable under moderate strain but increases noticeably upon entering the 10\% regime. Two key questions arise from this behavior: (i) can the apparent minor variations below 7\% be regarded as statistically negligible within experimental limits. (ii) how robust is the observed increase at 10\% when we consider two matters: first, we must ensure that the overall pooled effect is not being driven by one particular experimental subgroup,for example, that a single annealing or stacking condition does not disproportionately influence the 10\% increase. Second, the regression model used to quantify this effect should remain valid even if the data deviate from ideal statistical assumptions, such as having constant error variance or normally distributed residuals.\\
\subsection{Statistical Evaluation of Strain Sensitivity}
As mentioned earlier, the descriptive analysis establishes that band gap shifts remain minimal up to 7\% strain, while a clear increase emerges at 10\%. To quantify this pattern, we first fitted a pooled linear regression of 
\(\Delta E_g\) versus strain using only the points at 1--7\%. Here, ``pooled'' means that all samples across annealing and stacking conditions were combined into a single regression model. This provides a global estimate of the slope while ignoring subgroup structure. The estimated slope was 
\(-0.089~\mathrm{meV/\%}\) with a 90\% CI of \([-0.173, -0.005]~\mathrm{meV/\%}\). We then applied a two one-sided test (TOST) for equivalence, using a pre-specified margin of 
\(\pm 0.5~\mathrm{meV/\%}\). This margin was chosen as a conservative tolerance for practically negligible trends in \(\Delta E_g\). The TOST yielded an extremely small p-value ($p < 10^{-14}$), providing strong evidence that the slope is statistically equivalent to zero. Thus, it can be concluded that the optical band gap remains effectively invariant below 7\% strain.\\
To verify the robustness of the 10\% shift, we re-expressed strain as a categorical factor and fitted a regression with heteroskedasticity-consistent (HC3) standard errors. This approach avoids relying on the usual regression assumption of normally distributed and homoscedastic errors, and it reduces the influence of unusual or outlying samples by widening their estimated uncertainty so they cannot dominate the overall result. The coefficient for the 10\% condition was estimated at 
\(+4.27~\mathrm{meV}\) (95\% CI: +2.38 to +6.17~meV, \(p=9.4 \times 10^{-6}\)). These values confirm the statistically decisive increase even under conservative assumptions. Table \ref{tab:strain_tests} summarizes these data.\\
Although each subgroup independently showed a positive 10\% effect, to ensure that the pooled statistical contrast was not dominated by any single condition, we conducted leave-one-group-out sensitivity tests. Here, the overall 10\% coefficient was re-estimated while systematically excluding one anneal/stack combination at a time. The resulting values ranged from 3.66 to 4.80~meV, and in all cases the 95\% CIs remained strictly positive. This demonstrates that the band gap widening at 10\% strain is a universal feature across annealing and stacking variations, not a subgroup-specific anomaly.\\
To examine whether the magnitude of the 10\% effect depends on preparation or stack layers, we restricted the dataset to the 10\% condition and fitted a categorical regression with annealing temperature and stack type as predictors. The results showed neither annealing temperature nor stacking configuration contributed significantly to the variance in \(\Delta E_g\) (all \(p > 0.2\), HC3 regression). Nevertheless, inspection of the group means suggests that stacks containing a PEDOT:PSS interlayer tend to exhibit a somewhat larger shift compared to bare PET/P3HT. This trend is consistent with more efficient and uniform strain transfer into the P3HT layer when the adhesive layer is present, which was indeed the intended purpose of introducing PEDOT:PSS.\\
Finally, to test whether the 10\% increase merely reflects a continuation of the linear slope observed below 7\%, we fitted a pooled model of the form 
\[
\Delta E_g \sim \mathrm{strain} + \mathrm{is10},
\]
where \(\mathrm{is10}\) is a binary indicator for the 10\% condition. This formulation separates the effect into a baseline linear trend and any additional changes specific to 10\%. The deviation coefficient was estimated at 
\(+4.81~\mathrm{meV}\) (SE = 1.03, \(p = 3.1 \times 10^{-6}\)). This confirms that the 10\% regime exhibits a distinct upward departure from the pooled linear trend.\\
Together, these results demonstrate that the optical band gap of P3HT films remains stable up to 7\% strain, but exhibits a robust and reproducible widening of approximately 4--5~meV at 10\%. This effect is consistent across annealing and stacking variations and cannot be explained by extrapolation of the sub-threshold trend. Moreover, the somewhat stronger response in PEDOT:PSS-containing stacks suggests that the adhesive layer facilitates more effective strain transfer into the active P3HT, aligning with its design purpose.

\subsection{Physical Interpretation and Device-Level Implications}

The negligible band gap shifts observed up to 7\% strain suggest that the optoelectronic properties of P3HT films remain stable because strain is dissipated through intrinsic microstructural relaxation processes rather than electronic perturbation. Poly(3-hexylthiophene) is known to form semicrystalline films comprising ordered lamellae interspersed within amorphous matrices~\cite{Prosa1992,Ma2005}. These crystalline domains, stabilized by $\pi$–$\pi$ stacking and lamellar alignment, have been shown to provide mechanical reinforcement and preserve charge transport properties under moderate strain~\cite{Fan2020,Hollingsworth2023}. Strain in such films is accommodated through chain slippage or relaxation at phase boundaries \cite{Yu2023}. In addition, conformational reorganization within amorphous regions can occur, rather than direct deformation of the conjugated backbone. These mechanisms help maintain orbital overlap and prevent significant disruption of the electronic structure.\\
At higher strain levels (10\%), the modest but statistically significant increase in optical band gap ($\Delta E_g \approx 4.5$ meV) likely reflects a structural perturbation beyond the deformation tolerance of these buffering mechanisms. High-magnitude tensile strain can induce torsional disorder, chain misalignment, or disruption of favorable stacking orientations~\cite{Hollingsworth2023}. These structural changes reduce the effective conjugation length and lead to a blue-shift of the absorption onset. This interpretation is consistent with ultrafast microscopy results from Kim \textit{et al.}~\cite{Kim2021}, which showed that localized tensile strain suppresses $\pi$ orbital overlap and transiently increases $E_g$ within hundreds of picoseconds.\\
Importantly, the observed band gap modulation is largely independent of annealing condition or stack architecture. This suggests that the strain response is governed primarily by the semicrystalline network itself rather than interfacial coupling or surface chemistry. Moreover, since strain was applied and held for 30 minutes before UV–Vis remeasurement, the reported spectral shifts represent residual strain effects, reflecting post-deformation structural relaxation rather than transient or in-situ deformation.\\
\textbf{Bridging Molecular Simulations and Thin-Film Realities.}
While molecular-scale studies have elucidated fundamental strain–electronic coupling mechanisms, their results do not necessarily extrapolate to thin-film systems under real-world conditions. For instance, Menichetti \textit{et al.}~\cite{Menichetti2017} used DFT to show that strain can shift the donor and acceptor levels at the P3HT/PCBM interface, modulating energy offsets relevant for exciton dissociation. Similarly, Kim \textit{et al.}~\cite{Kim2021} used scanning ultrafast electron microscopy to capture transient strain-induced modulation of $E_g$ in P3HT with sub-nanosecond temporal resolution. Both studies provide invaluable mechanistic insights at the molecular or nanoscale level.\\
However, thin-film organic semiconductors exhibit complex and heterogeneous morphologies. Their structure is not uniform, meaning semi-crystalline domains are interspersed within amorphous matrices, and the interfaces often show significant roughness. Grain boundaries further add to this heterogeneity~\cite{Fan2020,Yu2023}. Mechanical deformation in semicrystalline polymer films is highly non-affine. This means that when the sample is stretched, the molecules do not move in lockstep with the overall strain field. Instead, their local displacements vary with chain orientation and microstructure and stress is redistributed through mesoscale relaxation pathways. For example, chains in amorphous regions can disentangle and slip past one another, crystalline–amorphous phase boundaries can shift or reorient, and grain interfaces may accommodate strain through frictional sliding. These processes dissipate mechanical energy without requiring proportional stretching of the $\pi$-conjugated backbones, thereby preserving orbital overlap and mitigating direct electronic disruption \cite{Fan2020,Gallant2020}.\\
Our findings suggests that such morphologies provide a degree of electronic resilience, suppressing band gap modulation even under residual strains of up to 10\%. Strain can alter orbital energies at the local scale. While strain may perturb orbital energies locally, the film’s ensemble-averaged optical response remains stable. This disconnect underscores the need to validate molecular-level predictions in the geometry of complete devices. In practice, molecular simulations and ultrafast imaging only gain full meaning when viewed within the structural and mechanical framework of thin-film architectures.

\section{Conclusion}

Our results demonstrate that P3HT-based thin films, both as single layers (PET/P3HT) and in PET/PEDOT:PSS/P3HT stacks, retain stable optical band gaps under tensile strain up to 7\%, with no statistically significant change. This invariance holds across annealing conditions, indicating that the semicrystalline morphology of P3HT allows internal relaxation of strain without perturbing the electronic structure.\\
However, spectra recorded after applying 10\% strain for 30 minutes revealed a consistent increase in $E_g$ of approximately $\sim$4--5~meV. This would suggest a threshold above which torsional disorder or disrupted $\pi$--$\pi$ stacking begins to alter the electronic structure. These findings bridge molecular predictions with thin-film realities, showing that thin films do not exhibit continuous strain sensitivity as simulations might suggest, but rather a plateaued response up to a critical strain level.\\
For device simulation including finite element modeling (FEM) of flexible optoelectronics, this behavior allows simplification so that the optical band gap can be treated as a constant parameter up to 7\% strain. This enables more accurate and efficient models focused on mechanical propagation and charge transport, rather than strain-modulated absorption.\\
In general, our study provides a quantitative strain threshold for optical property stability in P3HT films, which supports their viability in long-term flexible device applications and informs design rules for strain-resilient organic photovoltaics.

\begin{IEEEbiography}[{\includegraphics[width=1in,height=1.25in,clip,keepaspectratio]{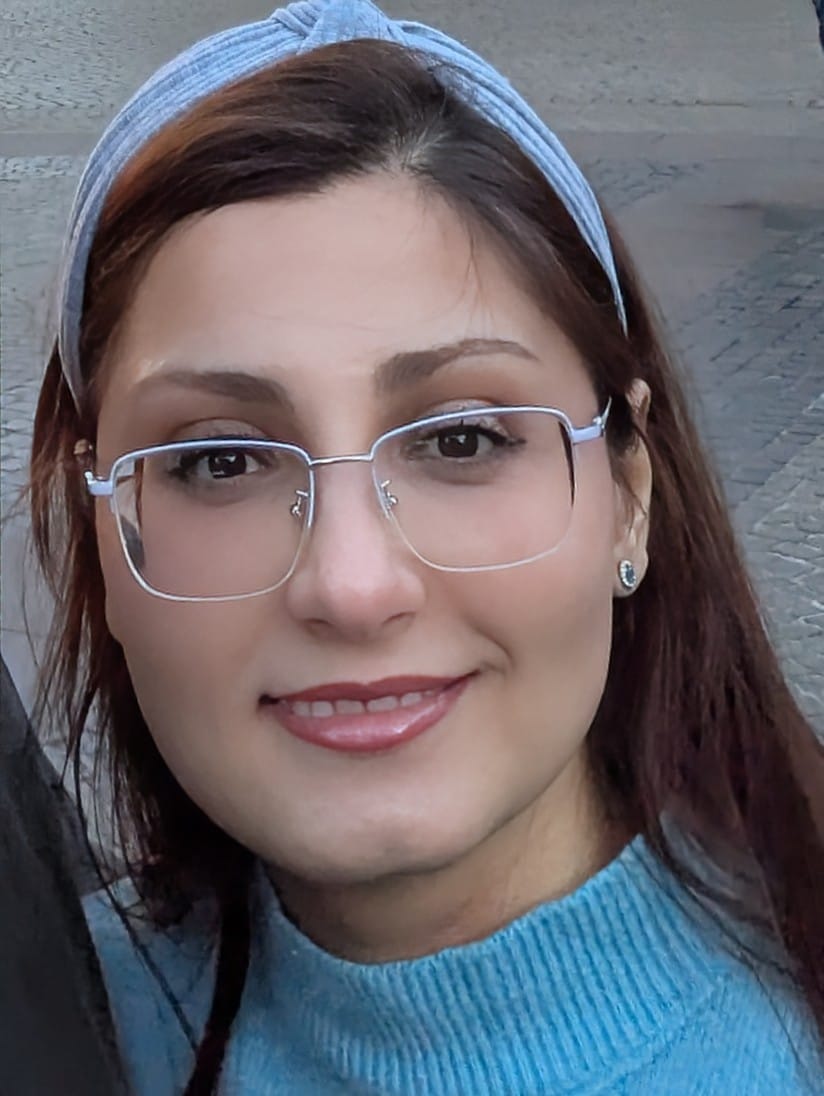}}]{Mahya Ghorab}
received the B.Sc. degree in electrical engineering from Ferdowsi University of Mashhad, Mashhad, Iran, in 2017, and the M.Sc. degree in micro and nano electronics from Shahrood University of Technology, Shahrood, Iran, in 2021, where she graduated first in her class. She is currently pursuing the Ph.D. degree in electrical engineering with Constructor University (formerly Jacobs University), Bremen, Germany, under the supervision of Prof.~Dr.-Ing.~Mojtaba Joodaki at the School of Computer Science and Electrical Engineering.
Her research focuses on the effect of mechanical strain on the electrical characteristics of organic solar cells, with an emphasis on understanding structure–property relationships relevant to flexible optoelectronic devices.
Ms.~Ghorab has authored and coauthored scientific papers in journals such as the \emph{IEEE Journal of Photovoltaics} and \emph{Advanced Photonics Research}, and has presented her work at international conferences on nanotechnology and materials science.
\end{IEEEbiography}
\vspace{-12em}
\begin{IEEEbiography}[{\includegraphics[width=1in,height=1.25in,clip,keepaspectratio]{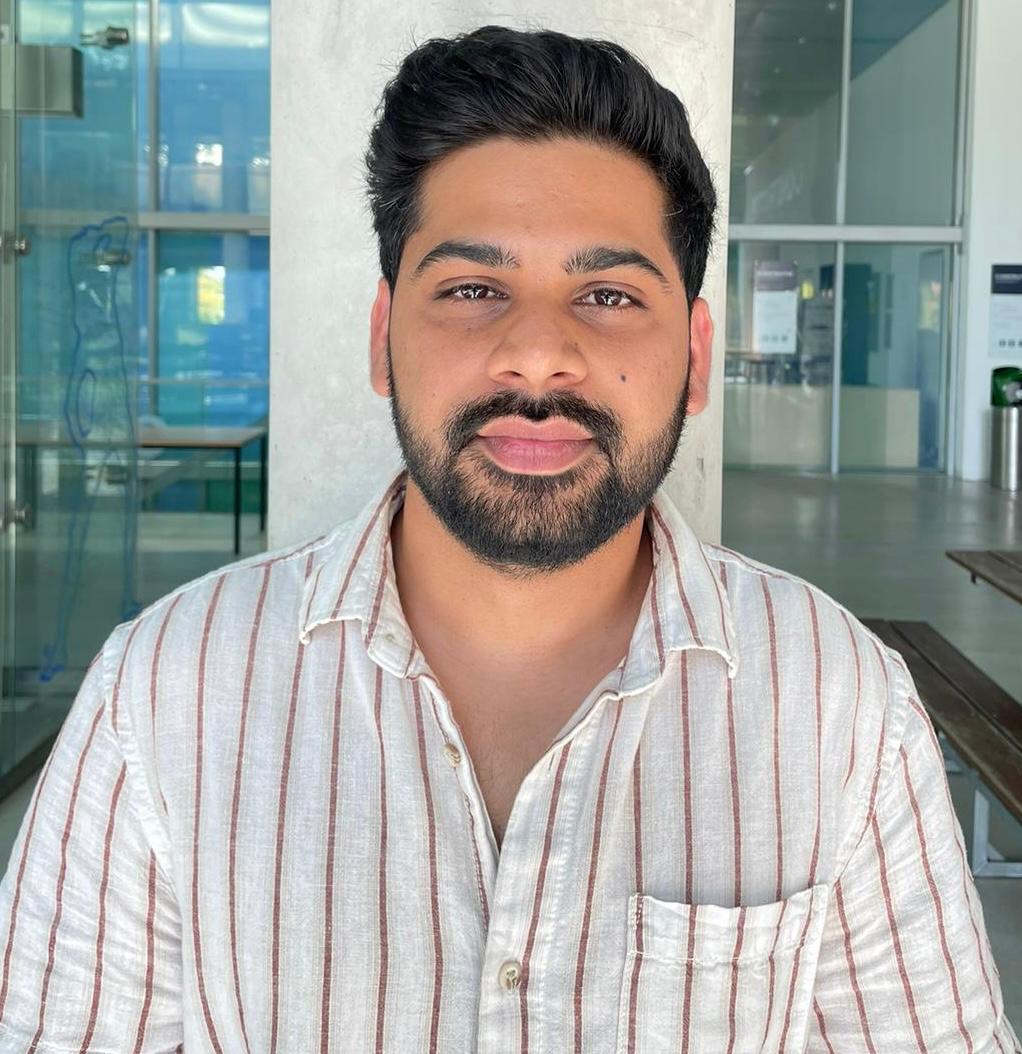}}]{Ayush Kant Ranga}
received the B.Sc. degree in physics from the University of Delhi, Delhi, India, in 2017, and the M.Sc. degree in physics from the Indian Institute of Technology, Gandhinagar, India, in 2020. He is currently pursuing the Ph.D. degree in physics with Constructor University (formerly Jacobs University), Bremen, Germany, under the supervision of Prof.~Arnulf Materny.
Since 2022, he has been working in the field of ultrafast time-resolved and Raman spectroscopy. His research interests focus on applying these techniques to investigate the fundamental processes in organic electronic materials.
Mr.~Ranga has coauthored scientific articles published in \emph{Physical Chemistry Chemical Physics} and \emph{Advanced Materials}, and has presented his research at international conferences in the fields of spectroscopy and photonics.
\end{IEEEbiography}
\begin{IEEEbiography}[{\includegraphics[width=1in,height=1.25in,clip,keepaspectratio]{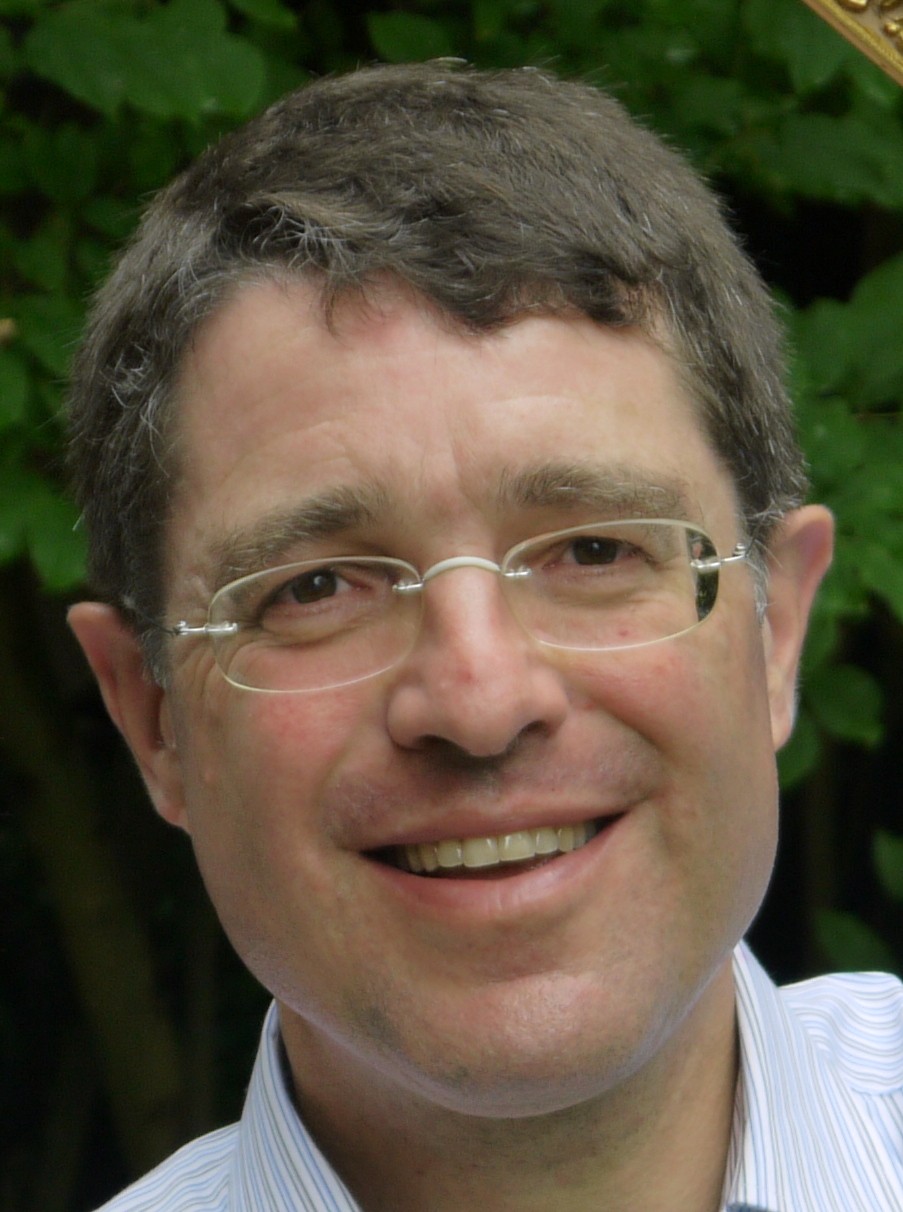}}]{Arnulf Materny}
received the Diploma degree in physics from the University of Bayreuth, Bayreuth, Germany, in 1988, where he graduated \emph{with distinction}, and the Dr.\ rer.\ nat.\ degree in chemistry and pharmacy from the University of Würzburg, Würzburg, Germany, in 1992, with \emph{summa cum laude} honors. He subsequently pursued postdoctoral and habilitation research with the research group of Nobel Laureate Prof.\ Ahmed Zewail at the California Institute of Technology, Pasadena, CA, USA, and with Prof.\ Wolfgang Kiefer at the University of Würzburg. He obtained the Dr.\ rer.\ nat.\ habil.\ degree in 1998 for his work on time-resolved optical spectroscopy of photoinduced intramolecular dynamics in isolated and environment-coupled molecules.\\
He was a Heisenberg Fellow and \emph{Privatdozent} at the University of Würzburg from 1999 to 2001. Since 2001, he has been a Full Professor of Chemical Physics at the School of Science, Constructor University (formerly International University Bremen / Jacobs University Bremen), Bremen, Germany.\\
Prof.~Materny’s research focuses on the techniques and applications of laser spectroscopy in both frequency and time domains. His work covers ultrafast molecular dynamics—including charge generation and loss in semiconductors, energy transfer in ionic liquids, and optimal control of chemical reactions using shaped femtosecond laser pulses—as well as the combination of nanometer spatial and femtosecond temporal resolution. He also specializes in linear and nonlinear Raman spectroscopy, with applications in organic semiconductor characterization, biofilm formation on microplastics, and industrial collaborative research.\\
He has authored and coauthored numerous scientific papers in journals such as \emph{The Journal of Physical Chemistry Letters}, \emph{ChemPhysChem}, \emph{Spectrochimica Acta Part A}, \emph{The Journal of Raman Spectroscopy}, \emph{Nanophotonics}, and \emph{Materials Today Communications}.\\
Prof.~Materny has been the recipient of several honors, including the Hoechst Research Award, the Kekulé Scholarship from the VW Foundation and the Chemical Industry Fund, and the Heisenberg Fellowship of the German Research Foundation (DFG). He is a Fellow of the World Innovation Foundation and serves as a long-standing member of multiple professional and academic committees, including the German Bunsen Society and the Committee for European Nonlinear Optical Spectroscopy.\\
He has served on the Editorial Boards of the \emph{Journal of Raman Spectroscopy} (since 2000) and \emph{Asian Chemical Letters} (since 2007) and continues to contribute actively to the international scientific community through his involvement in research collaborations and doctoral mentoring programs.
\end{IEEEbiography}

\begin{IEEEbiography}[{\includegraphics[width=1in,height=1.25in,clip,keepaspectratio]{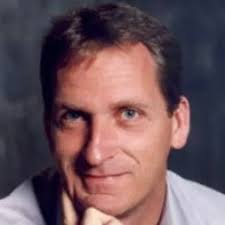}}]{Veit Wagner}
received the Diploma degree in physics from RWTH Aachen University, Aachen, Germany, in 1991, and the Ph.D. degree in physics from the Technical University of Berlin and RWTH Aachen University, Germany, in 1995. He completed his Habilitation in physics at the University of Würzburg, Würzburg, Germany, in 2001.\\
From 1999 to 2002, he was an Assistant Professor of Physics at the University of Würzburg. Since 2002, he has been a Professor of Physics at the School of Science, Constructor University (formerly Jacobs University), Bremen, Germany.\\
Prof.~Wagner’s research interests lie in molecular and nanoelectronics, (hybrid) organic systems, surface and interface properties, 2D materials, and novel materials for electronic applications, with emphasis on the development and characterization of thin-film devices, electronics on flexible substrates, and charge-carrier mobility and stability in organic and hybrid systems. He has authored and coauthored numerous journal articles, including works in \emph{Advanced Materials}, \emph{Applied Physics Letters}, and \emph{Organic Electronics}.\\
Prof.~Wagner is a member of the Deutsche Physikalische Gesellschaft (DPG) and the Materials Research Society (MRS, USA). He received the Borchers-Plakette of RWTH Aachen University in 1995 and serves on professional committees at Constructor University and beyond.
\end{IEEEbiography}

\begin{IEEEbiography}[{\includegraphics[width=1in,height=1.25in,clip,keepaspectratio]{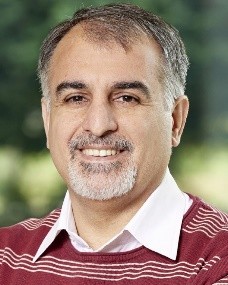}}]{Mojtaba Joodaki}
(Senior Member, IEEE) received the Ph.D. degree in electrical engineering from Kassel University, Kassel, Germany, in 2002. In the same year, he joined ATMEL Germany GmbH, Heilbronn, Germany, as an R\&D Engineer, where he worked on RF Si- and SiGe-based devices and circuits. In April 2005, he joined Infineon Technologies AG, Munich, Germany, as a Development Engineer, where he was responsible for EMI/EMC aspects of memory modules. From October 2006 to July 2009, he was a Device Engineer with Qimonda GmbH, Dresden, Germany, involved in the development of nanotransistors for dynamic random-access memory (DRAM) products.\\
He subsequently joined the Institute of Nanostructure Technologies and Analytics (INA), Kassel University, as a Visiting Scientist and Lecturer, where he defended his Habilitation dissertation in April 2011. In 2010, he joined the Department of Electrical Engineering, Ferdowsi University of Mashhad, Mashhad, Iran, as an Assistant Professor of Electronic Engineering (RF Circuit Design and Semiconductor Technology). He was promoted to Associate Professor in September 2011 and to Professor in July 2018. Since January 2020, he has been with Constructor University (formerly Jacobs University), Bremen, Germany, as a Professor of Electrical Engineering and Program Chair of Electrical and Computer Engineering.\\
His research interests include modeling, characterization, and fabrication of passive and active devices (organic and inorganic) for high-frequency, optoelectronic, and memory applications, as well as EMC of electronic products.\\
Prof.~Joodaki is a Life Member of the International Society for Optics and Photonics (SPIE). He has received several awards for his scientific achievements, including the Best Dissertation Prize of North Hessen Universities from the Association of German Engineers (VDI) in 2004, the Young Graduate Research Fellowship of the Gallium Arsenide (GaAs) Association at the European Microwave Week in 2001 and 2002, and the F-Made Scholarship of SPIE in 2002.
\end{IEEEbiography}

\EOD


\begin{thebibliography}{00}

\bibitem{WARDANI}
N.~K.~Wardani, M.~Jahandar, J.~Heo, Y.~H.~Kim, S.~Kim, and D.~C.~Lim, 
``Enhancing Physicochemical Durability and Photoelectronic Performance Beyond Bendable Large-Area Organic Photoelectronic Devices Through Tailored Multilayer Interface,'' 
Chemical Engineering Journal, vol.~502, p.~157958, 2024, doi: \href{https://doi.org/10.1016/j.cej.2024.157958}{10.1016/j.cej.2024.157958}.


\bibitem{Dauzon2021}
E.~Dauzon, X.~Sallenave, C.~Plesse, F.~Goubard, A.~Amassian, and T.~D.~Anthopoulos, 
``Pushing the Limits of Flexibility and Stretchability of Solar Cells: A Review,'' 
Advanced Materials, vol.~33, no.~36, p.~2101469, 2021, doi: \href{https://doi.org/10.1002/adma.202101469}{10.1002/adma.202101469}.

\bibitem{Wang2018}
G.~J.~N.~Wang, A.~Gasperini, and Z.~Bao, 
``Stretchable Polymer Semiconductors for Plastic Electronics,'' 
Advanced Electronic Materials, vol.~4, no.~2, p.~1700429, 2018, doi: \href{https://doi.org/10.1002/aelm.201700429}{10.1002/aelm.201700429}.

\bibitem{Ghorab2025}
M.~Ghorab, A.~K.~Ranga, P.~Donfack, A.~Materny, V.~Wagner, and M.~Joodaki, 
``Strain-Induced Optical and Molecular Transformations in Poly(Ethylene Terephthalate) Films for Organic Electronic Applications,'' 
Advanced Photonics Research, Art.~no.~2500081, 2025, doi: \href{https://doi.org/10.1002/adpr.202500081}{10.1002/adpr.202500081}.

\bibitem{Kim2021}
T.~Kim, S.~Oh, U.~Choudhry, C.~D.~Meinhart, M.~L.~Chabinyc, and B.~Liao, 
``Transient Strain-Induced Electronic Structure Modulation in a Semiconducting Polymer Imaged by Scanning Ultrafast Electron Microscopy,'' 
Nano Letters, vol.~21, no.~21, pp.~9146--9152, 2021, doi: \href{https://doi.org/10.1021/acs.nanolett.1c02963}{10.1021/acs.nanolett.1c02963}.

\bibitem{Menichetti2017}
G.~Menichetti, R.~Colle, and G.~Grosso, 
``Strain Modulation of Band Offsets at the PCBM/P3HT Heterointerface,'' 
The Journal of Physical Chemistry C, vol.~121, no.~25, pp.~13707--13716, 2017, doi: \href{https://doi.org/10.1021/acs.jpcc.7b02717}{10.1021/acs.jpcc.7b02717}.


\bibitem{Aboutorabi2015}
R.~Z.~Aboutorabi and M.~Joodaki, 
``Thermal Analysis of Organic Solar Cells Using an Enhanced Opto-Thermal Model,'' 
Organic Electronics, vol.~25, pp.~184--192, Oct. 2015, doi: \href{https://doi.org/10.1016/j.orgel.2015.06.034}{10.1016/j.orgel.2015.06.034}.

\bibitem{Aboutorabi2014}
R.~Z.~Aboutorabi and M.~Joodaki, 
``A Tripartite Physics-Based Model for Organic Solar Cells,'' 
in Proceedings of the 29th European Photovoltaic Solar Energy Conference and Exhibition (EU PVSEC 2014), Munich, Germany, Sep. 2014, pp.~1592--1596, doi: \href{https://doi.org/10.4229/EUPVSEC20142014-3BV.5.46}{10.4229/EUPVSEC20142014-3BV.5.46}.

\bibitem{Khorami2019}
A.~Khorami and M.~Joodaki, 
``Extracting Voltage-Dependent Series Resistance of Single Diode Model for Organic Solar Cells,'' 
SN Applied Sciences, vol.~1, no.~6, Art.~no.~619, Jun. 2019, doi: \href{https://doi.org/10.1007/s42452-019-0613-2}{10.1007/s42452-019-0613-2}.



\bibitem{Ghorab2022}
M.~Ghorab, A.~Fattah, and M.~Joodaki, 
``Tensile Mechanical Strain Effects on the Electrical Characteristics of Roll-to-Roll Printed OSC,'' 
IEEE Journal of Photovoltaics, vol.~12, no.~3, pp.~737--743, 2022, doi: \href{https://doi.org/10.1109/JPHOTOV.2022.3148717}{10.1109/JPHOTOV.2022.3148717}.

\bibitem{Joodaki2018}
M.~Joodaki and M.~Salari, 
``Investigation of the Tensile Strain Influence on Flicker Noise of Organic Solar Cells Under Dark Condition,'' 
Organic Electronics, vol.~59, pp.~230--235, 2018, doi: \href{https://doi.org/10.1016/j.orgel.2018.05.018}{10.1016/j.orgel.2018.05.018}.

\bibitem{Salari2018}
M.~Salari, M.~Joodaki, and S.~Mehregan, 
``Experimental Investigation of Tensile Mechanical Strain Influence on the Dark Current of Organic Solar Cells,'' 
Organic Electronics, vol.~54, pp.~192--196, 2018, doi: \href{https://doi.org/10.1016/j.orgel.2017.12.047}{10.1016/j.orgel.2017.12.047}.

\bibitem{Salari2019}
M.~Salari and M.~Joodaki, 
``Investigation of Electrical Characteristics Dependency of Roll-to-Roll Printed Solar Cells With Silver Electrodes on Mechanical Tensile Strain,'' 
IEEE Transactions on Device and Materials Reliability, vol.~19, no.~4, pp.~718--722, 2019, doi: \href{https://doi.org/10.1109/TDMR.2019.2949116}{10.1109/TDMR.2019.2949116}.


\bibitem{Sokolov2012}
A.~N.~Sokolov, Y.~Cao, O.~B.~Johnson, and Z.~Bao, 
``Mechanistic Considerations of Bending-Strain Effects Within Organic Semiconductors on Polymer Dielectrics,'' 
Advanced Functional Materials, vol.~22, no.~1, pp.~175--183, 2012, doi: \href{https://doi.org/10.1002/adfm.201101418}{10.1002/adfm.201101418}.

\bibitem{Wu2016}
Y.~Wu, A.~R.~Chew, G.~A.~Rojas, G.~Sini, G.~Haugstad, A.~Belianinov, S.~V.~Kalinin, H.~Li, C.~Risko, J.-L.~Brédas, A.~Salleo, and C.~D.~Frisbie, 
``Strain Effects on the Work Function of an Organic Semiconductor,'' 
Nature Communications, vol.~7, Art.~no.~10270, 2016, doi: \href{https://doi.org/10.1038/ncomms10270}{10.1038/ncomms10270}.

\bibitem{Kubo2016}
T.~Kubo, R.~Häusermann, J.~Tsurumi, J.~Soeda, Y.~Okada, Y.~Yamashita, N.~Akamatsu, A.~Shishido, C.~Mitsui, T.~Okamoto, S.~Yanagisawa, H.~Matsui, and J.~Takeya, 
``Suppressing Molecular Vibrations in Organic Semiconductors by Inducing Strain,'' 
Nature Communications, vol.~7, Art.~no.~11156, 2016, doi: \href{https://doi.org/10.1038/ncomms11156}{10.1038/ncomms11156}.

\bibitem{Ghorabreview}
M.~Ghorab, A.~Fattah, and M.~Joodaki, 
``Fundamentals of Organic Solar Cells: A Review on Mobility Issues and Measurement Methods,'' 
Optik, vol.~267, Art.~no.~169730, 2022, doi: \href{https://doi.org/10.1016/j.ijleo.2022.169730}{10.1016/j.ijleo.2022.169730}.

\bibitem{Khoshdel2019}
V.~Khoshdel, M.~Joodaki, and M.~Shokooh-Saremi, 
``UV and IR Cut-Off Filters Based on Plasmonic Crossed-Shaped Nano-Antennas for Solar Cell Applications,'' 
Optics Communications, vol.~433, pp.~275--282, 2019, doi: \href{https://doi.org/10.1016/j.optcom.2018.10.005}{10.1016/j.optcom.2018.10.005}.

\bibitem{Saleh2021}
R.~Saleh, M.~Barth, W.~Eberhardt, and A.~Zimmermann, 
``Bending Setups for Reliability Investigation of Flexible Electronics,'' 
Micromachines, vol.~12, no.~1, p.~78, 2021, 
doi: \href{https://doi.org/10.3390/mi12010078}{10.3390/mi12010078}.


\bibitem{Huang2014}
H.~Huang and J.~Huang, Eds., 
Organic and Hybrid Solar Cells. 
Cham, Switzerland: Springer International Publishing, 2014, doi: \href{https://doi.org/10.1007/978-3-319-10855-1}{10.1007/978-3-319-10855-1}.

\bibitem{Bredas2009}
J.-L.~Brédas, J.~E.~Norton, J.~Cornil, and V.~Coropceanu, 
``Molecular Understanding of Organic Solar Cells: The Challenges,'' 
Accounts of Chemical Research, vol.~42, no.~11, pp.~1691--1699, 2009, doi: \href{https://doi.org/10.1021/ar900099h}{10.1021/ar900099h}.

\bibitem{Dimitriev2022}
O.~P.~Dimitriev, 
``Dynamics of Excitons in Conjugated Molecules and Organic Semiconductor Systems,'' 
Chemical Reviews, vol.~122, no.~9, pp.~8487--8593, 2022, doi: \href{https://doi.org/10.1021/acs.chemrev.1c00648}{10.1021/acs.chemrev.1c00648}.



\bibitem{Scharber2013}
M.~C.~Scharber and N.~S.~Sariciftci, 
``Efficiency of Bulk-Heterojunction Organic Solar Cells,'' 
Progress in Polymer Science, vol.~38, no.~12, pp.~1929--1940, 2013, doi: \href{https://doi.org/10.1016/j.progpolymsci.2013.05.001}{10.1016/j.progpolymsci.2013.05.001}.
\bibitem{Pittner2013}
S.~Pittner, D.~Lehmann, D.~R.~T.~Zahn, and V.~Wagner, 
``Charge Transport Analysis of Poly(3-hexylthiophene) by Electroreflectance Spectroscopy,'' 
Physical Review B, vol.~87, no.~11, p.~115211, Mar. 2013, 
doi: \href{https://doi.org/10.1103/PhysRevB.87.115211}{10.1103/PhysRevB.87.115211}.


\bibitem{Spano2010}
F.~C.~Spano, 
``The Spectral Signatures of Frenkel Polarons in H- and J-Aggregates,'' 
Accounts of Chemical Research, vol.~43, no.~3, pp.~429--439, 2010, doi: \href{https://doi.org/10.1021/ar900233v}{10.1021/ar900233v}.

\bibitem{Tauc1966}
J.~Tauc, R.~Grigorovici, and A.~Vancu, 
``Optical Properties and Electronic Structure of Amorphous Germanium,'' 
physica status solidi (b), vol.~15, no.~2, pp.~627--637, 1966, doi: \href{https://doi.org/10.1002/pssb.19660150224}{10.1002/pssb.19660150224}.

\bibitem{Yu2010}
P.~Y.~Yu and M.~Cardona, 
Fundamentals of Semiconductors: Physics and Materials Properties, 4th~ed. 
Berlin, Heidelberg: Springer-Verlag, 2010, doi: \href{https://doi.org/10.1007/978-3-642-00710-1}{10.1007/978-3-642-00710-1}.

\bibitem{Kerremans2020}
R.~Kerremans, C.~Kaiser, W.~Li, N.~Zarrabi, P.~Meredith, and A.~Armin, 
``The Optical Constants of Solution-Processed Semiconductors—New Challenges With Perovskites and Non-Fullerene Acceptors,'' 
Advanced Optical Materials, vol.~8, no.~16, Art.~no.~2000319, 2020, doi: \href{https://doi.org/10.1002/adom.202000319}{10.1002/adom.202000319}.


\bibitem{Zhong2023}
H.~Zhong, F.~Pan, S.~Yue, C.~Qin, V.~Hadjiev, F.~Tian, X.~Liu, F.~Lin, Z.~Wang, and J.~Bao, 
``Idealizing Tauc Plot for Accurate Bandgap Determination of Semiconductor With Ultraviolet--Visible Spectroscopy: A Case Study for Cubic Boron Arsenide,'' 
The Journal of Physical Chemistry Letters, vol.~14, no.~29, pp.~6702--6708, 2023, doi: \href{https://doi.org/10.1021/acs.jpclett.3c01416}{10.1021/acs.jpclett.3c01416}.

\bibitem{Scharber2021}
M.~C.~Scharber and N.~S.~Sariciftci, 
``Low Band Gap Conjugated Semiconducting Polymers,'' 
Advanced Materials Technologies, vol.~6, no.~4, Art.~no.~2000857, 2021, doi: \href{https://doi.org/10.1002/admt.202000857}{10.1002/admt.202000857}.

\bibitem{Makula2018}
P.~Makuła, M.~Pacia, and W.~Macyk, 
``How to Correctly Determine the Band Gap Energy of Modified Semiconductor Photocatalysts Based on UV--Vis Spectra,'' 
The Journal of Physical Chemistry Letters, vol.~9, no.~23, pp.~6814--6817, 2018, doi: \href{https://doi.org/10.1021/acs.jpclett.8b02892}{10.1021/acs.jpclett.8b02892}.

\bibitem{Hang2023}
P.~Hang, C.~Kan, B.~Li, Y.~Yao, Z.~Hu, Y.~Zhang, J.~Xie, Y.~Wang, D.~Yang, and X.~Yu, 
``Highly Efficient and Stable Wide-Bandgap Perovskite Solar Cells via Strain Management,'' 
Advanced Functional Materials, vol.~33, no.~11, Art.~no.~2214381, 2023, doi: \href{https://doi.org/10.1002/adfm.202214381}{10.1002/adfm.202214381}.

\bibitem{Mamand2025}
D.~M.~Mamand, D.~S.~Muhammad, D.~Q.~Muheddin, \emph{et al.}, 
``Optical Band Gap Modulation in Functionalized Chitosan Biopolymer Hybrids Using Absorption and Derivative Spectrum Fitting Methods: A Spectroscopic Analysis,'' 
Scientific Reports, vol.~15, Art.~no.~3162, 2025, doi: \href{https://doi.org/10.1038/s41598-025-87353-5}{10.1038/s41598-025-87353-5}.



\bibitem{Muller2017}
A.~Müller, V.~Jovanov, and V.~Wagner, 
``Analytical Model (CELIC) for Describing Organic and Inorganic Solar Cells Based on Drift-Diffusion Calculations,'' 
Applied Physics Letters, vol.~111, no.~2, Art.~no.~023506, Jul. 2017, doi: \href{https://doi.org/10.1063/1.4993778}{10.1063/1.4993778}.
\bibitem{Tsumuraya2012}
T.~Tsumuraya, J.-H.~Song, and A.~J.~Freeman, 
``Linear Optical Properties and Electronic structures of Poly(3-hexylthiophene) and Poly(3-hexylselenophene) Crystals from First Principles,'' 
Physical Review B, vol.~86, no.~7, p.~075114, Aug. 2012, 
doi: \href{https://doi.org/10.1103/PhysRevB.86.075114}{10.1103/PhysRevB.86.075114}.



\bibitem{Scott1992}
D.~W.~Scott, 
``Kernel Density Estimators,'' 
in Multivariate Density Estimation, 
ch.~6, pp.~125--193, John Wiley \& Sons, Ltd, 1992, doi: \href{https://doi.org/10.1002/9780470316849.ch6}{10.1002/9780470316849.ch6}.

\bibitem{Prosa1992}
T.~J.~Prosa, M.~J.~Winokur, J.~Moulton, P.~Smith, and A.~J.~Heeger, 
``X-Ray Structural Studies of Poly(3-alkylthiophenes): An Example of an Inverse Comb,'' 
Macromolecules, vol.~25, no.~17, pp.~4364--4372, 1992, doi: \href{https://doi.org/10.1021/ma00043a019}{10.1021/ma00043a019}.

\bibitem{Verploegen2010}
E.~Verploegen, R.~Mondal, C.~J.~Bettinger, S.~Sok, M.~F.~Toney, and Z.~Bao,  
``Effects of Thermal Annealing Upon the Morphology of Polymer–Fullerene Blends,''  
Advanced Functional Materials, vol.~20, no.~20, pp.~3519--3529, 2010,  
doi: \href{https://doi.org/10.1002/adfm.201000975}{10.1002/adfm.201000975}.


\bibitem{Ma2005}
W.~Ma, C.~Yang, X.~Gong, K.~Lee, and A.~J.~Heeger, ``Thermally Stable, Efficient
Polymer Solar Cells With Nanoscale Control of the Interpenetrating
Network Morphology,'' Advanced Functional Materials, vol.~15, no.~10,
pp.~1617--1622, 2005, doi: \href{https://doi.org/10.1002/adfm.200500211}{10.1002/adfm.200500211}.




\bibitem{Fan2020}
Q.~Fan, W.~Su, S.~Chen, T.-S.~Kim, L.~Huang, C.~Yang, and E.~Wang, 
``Mechanically Robust All-Polymer Solar Cells from Narrow Band Gap Acceptors with Hetero-Bridging Atoms,'' 
Joule, vol.~4, pp.~658--672, 2020, doi: \href{https://doi.org/10.1016/j.joule.2020.01.014}{10.1016/j.joule.2020.01.014}.

\bibitem{Hollingsworth2023}
N.~Hollingsworth, S.~T. Hoffmann, and J.~Clark, 
``Strain-Dependent Photophysics in Conjugated Polymer Films: A Molecular Origin of Mechanical–Optical Coupling,'' 
The Journal of Physical Chemistry B, vol.~127, no.~9, pp.~2277--2285, 2023, doi: \href{https://doi.org/10.1021/acs.jpcb.3c00152}{10.1021/acs.jpcb.3c00152}.

\bibitem{Yu2023}
H.~Yu, Y.~Wang, X.~Zou, J.~Yin, X.~Shi, Y.~Li, H.~Zhao, L.~Wang, H.~M.~Ng, B.~Zou, X.~Lu, K.~S.~Wong, W.~Ma, Z.~Zhu, H.~Yan, and S.~Chen, 
``Improved Photovoltaic Performance and Robustness of All-Polymer Solar Cells Enabled by a Polyfullerene Guest Acceptor,'' 
Nature Communications, vol.~14, art.~2323, 2023, doi: \href{https://doi.org/10.1038/s41467-023-37738-9}{10.1038/s41467-023-37738-9}.

\bibitem{Gallant2020}
B.~M. Gallant, H.~Kang, and A.~Amassian, 
``Highly Stretchable and Mechanically Stable Organic Solar Cells Fabricated Using a Conductive Adhesive Electrode,'' 
Matter, vol.~2, no.~1, pp.~147--163, 2020, doi: \href{https://doi.org/10.1016/j.matt.2019.10.015}{10.1016/j.matt.2019.10.015}.














\end{thebibliography}
\end{document}